\documentstyle[preprint,aps,epsf]{revtex}

\begin{document}
\preprint{IASSNS-HEP-97/34}
\draft
\title{
Adaptation and Optimal Chemotactic Strategy for {\it E. Coli} } 
\author{
 S. P. Strong$^{(1,\clubsuit)}$,
B. Freedman$^{(2,\diamondsuit)}$,
William Bialek$^{(1)}$ and R. Koberle$^{(1,\spadesuit)}$
 }

\address{
$^{(1)}$NEC Research Institute, 4 Independence Way, Princeton, NJ 08540 \\
$^{(2)}$Department of Physics, Harvard University, Cambridge MA, 02138\\
}
\date{\today}
\maketitle

\begin{abstract}

Extending  the classic works
of Berg and Purcell on
the biophysics of  bacterial
chemotaxis,
we find the optimal chemotactic strategy for
the 
peritrichous bacterium {\it E. Coli}
in the high and low signal 
to noise ratio limits.
The optimal strategy depends 
on properties of the environment and
properties of the individual bacterium and
is therefore highly adaptive. We review 
experiments relevant to testing both the
form of the proposed strategy and its
adaptability and propose extensions
of them which could test the limits of
the adaptability in this 
simplest sensory processing system.

\vspace{.75in}

\centerline{$^{\clubsuit}$present address:  Institute for Advanced Study,
Olden Lane, Princeton NJ, 08540 }
\centerline{$^{\diamondsuit}$present address:  William M. Mercer Inc.,
1166 Avenue of the Americas, New York NY, 10036 }
\centerline{$^{\spadesuit}$present address:  Instituto de F\'isica de S\~ao
Carlos, Universidade de S\~ao Paolo,}
\centerline{C. Postal 369, 13560-250 S\~ao
Carlos SP Brasil}

\end{abstract}

\subsection{Introduction}
If placed in an inhomogeneous solution of a chemoattractant
such as $\alpha$-methyl-D,L-aspartate, {\it E. Coli}
collect visibly in the regions of high concentration
of the attractant
\cite{receptors,Breview1}.   There is a parallel phenomenon
that occurs for chemorepellants
where the bacteria collect in regions
of low repellant concentration.
The two phenomena are referred to as chemotaxis
and have been known in a variety of
bacteria since the 1880's 
\cite{oldlit}.  They constitute perhaps the
simplest known example of 
sensory processing dependent behavior 
in a living organism.
Actual sensory processing must be involved since,
for  {\it E. Coli}, the chemoattractant need not
be a substance that the bacterium can metabolize
in any way
and it is known that the response
to the chemoattractant relies on 
specific chemoreceptors on the outer
membrane of the bacterium which 
bind chemoattractant and signal
their occupancy to the interior of the cell through a
phosphorylation cascade \cite{receptors,phosphorcascade}. 
The manner in which the bacteria
use this information to
reach the regions of high
chemoattractant concentration
was first illuminated by
Berg and Brown \cite{Btracking}
who built a special microscope to track the
motions of the individual bacteria.
What they saw were stretches of motion at approximately
$10\mu$m/sec with a slowly varying direction
of orientation 
(termed ``runs'')
separated by periods
when the bacterium came to a stop and
changed orientation very rapidly 
(referred to as  ``tumbles'') before 
continuing on in another run.
We now know that both of these
characteristic motions of {\it E. Coli},
and ultimately chemotaxis, are 
due to the  rotation of the flagella
\cite{PlowR,bacteriaswim,lowRfootnote,rotationchemotact}.
{\it E. Coli} typically have several
flagella spread over their surfaces
(this is what is meant
by the designation peritrichous)
and runs result from 
counterclockwise rotation of all the flagella.
For that direction of rotation,
the flagella come together to form a ``bundle''
and cooperate to  propel the bacterium.
Because the flagella are helical, the opposite
sense of rotation has quite a different effect:
the flagellar bundle comes apart, the bacterium
is not propelled and its orientation 
varies rapidly, but apparently randomly \cite{Btracking},
resulting in a tumble. 
Chemotaxis results from the coordination of
the tumbling times with the time dependence of
the receptor occupancies so that the bacteria
change direction less often when
they are headed in the direction of increasing
chemoattractant \cite{tempcomp1,rotationchemotact}.
The problem we will
discuss in this work is the optimal
strategy for coordinating these tumbles
with the input from the chemoreceptors.

Clearly, for an optimal strategy to exist
at all a problem must be very highly
constrained.  The principle constraints
which make this problem solvable are 
taken largely from the  experimental literature
on chemotaxis and motility of {\it E. Coli}.
First, as was pointed out by 
Berg and Purcell \cite{BandPmainchemorecept} in this context
and as we will briefly discuss, due to
rotational Brownian motion, 
{\it E. Coli} 
can not maintain an orientation for
an extended period of time.
Second, {\it E. Coli}
make no controlled changes of direction. 
It is 
obvious, given that they can not maintain
orientation that,
for sensory processing reasons alone,
they are incapable 
of turning in a specific
direction \cite{BandPmainchemorecept,Btracking},
however a change of direction
of a controlled magnitude is in principle
possible.  In practice, there is some evidence that
the length of tumbles is affected by
sensory input under some
circumstances \cite{tempcomp2};
however, this is not believed to be
important for chemotaxis under realistic
conditions \cite{Btracking}. 
We therefore make the assumption that
{\it E. Coli} change direction
by entering into ``tumbles''
which have have no characteristics which depend
on sensory input.
It seems likely that, in view
of the limited use {\it E. Coli}
could make use of steering capability
given its orientation 
problems,
this simple method of direction change 
was evolutionarily preferred because the
cost associated with this
capability are lower than 
those that a more developed
steering capability would impose.
In any case, we assume the tumbles 
are all identical 
and effectively randomize the orientation
of the bacteria 
over the course of a time 
$\tau_{\rm tumble} \sim .15$ sec
\cite{tumblefootnote}.
Finally, we  assume that during runs
the bacterium swims with
fixed velocity, $v$, independent of
sensory inputs.  This is known to be
approximately true experimentally
\cite{Btracking}.

Taken together, these three constraints
are sufficient to allow us to determine
the optimal chemotactic strategy
in the high signal to noise
ratio limit for the following definition of 
optimality:
we call a chemotactic strategy optimal
if it maximizes the expectation
value of $\langle \vec{v} \cdot \nabla c \rangle$
for a static, uniform gradient.
Here $\vec{v}$ is the swimming direction of
the bacterium
and
the average is
over the time history of the bacterium's path.
There are 
alternative definitions
of optimality
such as that of maximizing
$\langle c \rangle$
or 
$\frac{\langle \vec{v} \cdot \nabla c \rangle_r^2}{\langle \vec{v}\rangle_r^2}$
(here the $r$ subscript denotes an average only
over the runs
\cite{Leonid}), but we expect them to result 
in very similar optimal strategies provided
they  do not incorporate directly
(1) game-theoretic competition between
bacteria or (2) a complex structure
of maxima and minima 
in the concentration so that
gradient descent approaches such as
we are proposing become trapped in local 
minima.  The neglect of the latter 
possibility seems
very reasonable for realistic
environments, however, considerations
like the former may well have played an
important role in the evolution of {\it E. Coli}.
For example,
the known pattern forming behavior
of {\it E. Coli}
\cite{pattern} demonstrates that the
problem of chemotaxis
has special properties in
presence of depletable
nutrient and large numbers of
other {\it E. Coli}.  In fact,
our choice of optimality is partly motivated
by
the consideration of competition.
Maximizing $\langle \vec{v} \cdot \nabla c \rangle$
for static, uniform gradients
chooses the strategy that
leads the  bacterium to attractant most rapidly,
which is probably 
evolutionarily preferable to one that
leads the bacterium there more slowly,
but then results in the bacterium staying
slightly closer to the region of
maximal concentration.

Before discussing the strategy we obtain,
let us first make a general remark on the
nature of any optimal strategy:
with the definition
of optimality we have chosen,
the strategy must consist of a
deterministic algorithm
for deciding when to tumble based on
the history of chemoreceptor occupancy.
For any given history, the bacterium's
expected,
future, average $\vec{v} \cdot \nabla c$
is either raised or lowered by 
initiating a tumble
at that particular moment
and if it is raised the bacterium should
tumble and if it is lowered
it should not.  The stochastic nature of
observed runs and tumbles should result
(for an optimal strategy) entirely from
the stochastic nature of the inputs
(chemoreceptor bindings), not from any
deliberate introduction of ``noise''
in the decision making on the part of the
bacterium.  In practice,
a totally deterministic strategy
at fixed input requires an
amplifier of arbitrary fidelity
and gain to allow the inputs
to drive the decision making apparatus
and is not realizable; however,
the phosphorylation cascade seems capable,
in practice, of providing sufficient
gain and fidelity that the tumbling
would be effectively deterministic.
A deterministic strategy
is, however,  in some conflict with the
two state model proposed in \cite{tempcomp2} on
experimental grounds, where the past history of
the receptor occupancies is taken to modulate
the rates with which the flagellar motors 
change their direction of rotation.
In that strategy,
it is not the states but the
the rates for transitions between
the states that are set
for each flagellar motor and
they are all set separately,
so that the only correlations between different flagella
are rate-rate correlations. 
In that case,
the strategy would actually be stochastic
even for fixed inputs to the chemoreceptors.
However, in the limit that the rate modulations
for going from running to tumbling are large
(the rate is either $\sim 0$ or $\sim \infty$)
 the two state model
has a strategy of the form we propose.

As mentioned above, in practice, a purely ``deterministic'' strategy
for switching between two states is impossible
and, to approach it, large variations in
the rate of the above type could be used.
The rate model 
is thus one possibility for 
a realistic process which approaches the
optimal strategy and may represent {\it E. Coli}'s
best effort to implement the optimal strategy.
However, it should be noted that 
the evidence for the modulated rate model is
not entirely conclusive.
In particular, as we will see in our discussion of
run and tumble statistics for the optimal strategies,
the exponential tails present in the durations
of run and tumble times are {\it not}
uniquely explained by the rate model,
as claimed in \cite{tempcomp2}.
Further, there is a pronounced advantage that a deterministic
strategy has when the problem of flagellar coordination
is considered \cite{flagcoord}.

Free swimming {\it E. Coli} in the absence of
gradients spend roughly a tenth of their time
tumbling \cite{Btracking} 
while tethered bacteria 
\cite{tethernote} with a single flagellum
rotate it clockwise nearly half of the time.
Clearly, if the typical bacterium has 
five flagella, there
must be significant conflict between the
flagella in the absence of any signal.
The formation of a coherent
flagellar bundle 80\% of the time requires
some sort of coordination. Notice
that no coordination can
result from the signals
to the flagella if the signals are
independent,
as the model of \cite{tempcomp2}
predicts they should be at small
signal to noise ratio, where the above numbers
apply. 
At least one mechanism for the required coordination
has been proposed
\cite{flagcoord} and it seems clear that some
explanation is required, unless the 
data from tethered bacteria are taken to be
unrepresentative.
If the flagella are coordinated then
much of the self-induced ``noise'' 
implied by the stochastic nature of
their individual biases
will be eliminated by
the pooling of their inputs: 
an effectively deterministic strategy
will result.  In fact,
any ``deterministic'' strategy would have to result
either from some sort of cooperativity effect 
(to reduce the noise inherent in the stochastic nature of 
the binding and unbinding of individual internal 
signaling molecules to receptors) or from sufficient
amplification of the input signal to drive very large
variations in the concentration of internal signaling
molecule.  Cooperativity is generally a more efficient
and robust solution. 

It could be supplied by a either coordination of the
different flagella and pooling of their signals,
cooperative signaling to the individual flagellum 
or both.  In fact, the protein FliM which is believed to be
responsible for translating the internal tumble signal
\cite{Welch}, coming from the 
phosphorylated form of the 
protein CheY \cite{Borkovich},
is believed to be present in many (about 100) copies
for each flagellar motor \cite{Francis}.  It is not known
how many of these copies actually bind CheY or how this
binding is transduced into a tumble signal, but there
is clearly great potential for cooperativity.
For example, if the tumbling/running switch were 
whether more or less than 50 molecules of phosphorylated
CheY were bound, then the behavior would be nearly deterministic
for even moderate variations in the concentration
of phosphorylated CheY.

In view of this, it is not
clear that there is much difference in practice
between a deterministic signal and stochastic signals to the
individual motors or, perhaps, to the many individual CheY binding cites
at each motor, since the results of pooling the stochastic signals
might be nearly deterministic.  In so far as there is a difference,
the deterministic strategy will lead to
better performance, but may be an expensive 
capacity for {\it E. Coli} to maintain.

We note that the ``response regulator'' model
proposed in 
\cite{respreg}, in which the
tumbles are induced by threshold crossing
some functional of chemoreceptor binding histories,
is of the 
appropriately deterministic 
type and our optimal strategy will
be a realization of this  strategy where we specify,
at high signal to noise ratio, the correct threshold
(with the functional being somewhat arbitrary).
At low signal to noise ratio, both
the correct threshold and functional of the
receptor histories will be determined, subject to
certain assumptions about 
the receptor correlations. We will see
that the statistics of runs and tumbles
resulting in both cases are
not inconsistent with the 
statistics observed experimentally,
contrary to the claim of \cite{tempcomp2}
that threshold crossing strategies,
as opposed to rate modulation strategies,
disagree with the data.

\subsection{High Signal to Noise Ratio}
\label{sec:hSNRstrat}

Let us now begin with the ``high'' signal to
noise ratio case,
defined by the bacterium being
able to measure the projection
of the gradient onto its swimming direction
in a time much shorter than any other 
relevant time scale, in particular the time
scale set by rotational diffusion.  
For an object undergoing
rotational diffusion
\begin{equation}
\langle \hat{n}(t) \cdot \hat n(0) \rangle = \exp(- 2 D_{\rm rot} t)
\end{equation}
where $D_{\rm rot}$ is the rotational diffusion
constant. For {\it E. Coli} it is know empirically to
be about $.15~{\rm radian}^2/{\rm sec}$,
implying a time scale for rotational
Brownian motion of about $3$ sec. 
This is in
rough agreement with what one expects for the
case of a sphere
of radius $r=1\mu$m in room-temperature water.
There
\begin{eqnarray}
D_{\rm rot} & \sim & k T /\nu_{\rm rot} \\
\nu_{\rm rot} & \sim &  8 \pi \eta r^3
\\  \nonumber & \sim &
2.5 ~10^{-13} {\rm cm^2/sec} \\
D_{\rm rot} & \sim & .16 ~{\rm radian}^2/{\rm sec} 
\end{eqnarray}
where $T$ is the temperature,
$\nu_{\rm rot}$ is the
rotational drag on the sphere, and $\eta$ is the viscosity of
water.
Note that,
since we are at low Reynolds
number,
the rotational diffusion which is disorienting
the bacterium is Markovian. 
Its present orientation embodies
{\it all} of its knowledge about the
future and there is no need to
keep track of an angular momentum.
Because all of the information
about the future is contained
in the present orientation,
or equivalently the
bacterium's present
knowledge of $\vec{v} \cdot \nabla c$
obtained from the chemoreceptor histories,
the optimal strategy 
in the high signal to noise ratio limit
requires only the specification
of a single number: 
a threshold, $T$,
which is the maximum value of
$\vec{v} \cdot \nabla c$
which the bacterium will tolerate
before tumbling.
Any processing strategy for the
chemoreceptor histories that allows
the determination of $\vec{v} \cdot \nabla c$
rapidly compared to other time scales
is acceptable.
This makes this
limit particularly simple to discuss.

To solve the high signal to noise ratio limit
we need to make the assumption that each tumble
is perfectly disorienting and totally randomizes 
the bacterium orientation.  This
is approximately true \cite{Btracking} and
allows a full solution of this case.
This is interesting
even though the bacterium may rarely,
if ever, encounter a high signal to noise ratio
environment because the optimal strategy can
be solved for completely and 
it gives us an idea what the low
signal to noise ratio strategy is
evolving towards as the signal to noise
ratio is increased.  In
this limit,
there is still an interesting strategy
because the bacterium can not steer,
only reorient itself through
stereotyped tumbles, each of which requires
a finite amount of time.  
This finite time cost will be very important 
in determining the optimal strategy and will
result in the bacterium displaying a surprisingly
large amount of ``optimism'', by which we
mean that bacteria which know they are not swimming
directly up the gradient should choose to continue running 
because of the chance that rotational diffusion will
improve their prospects faster than tumbling would.
 
Let us consider first the case where the bacterium is in
a uniform  gradient of chemoattractant
and therefore has had time to
measure the magnitude of $\nabla c$. In this
case, it can translate it's knowledge of
$\vec{v} \cdot \nabla c$ directly into
knowledge of $\Theta$, the angle between
its motion and the concentration gradient,
and the strategy consists of an angle at which
the bacterium ``decides'' it is moving sufficiently in
the wrong direction that it is worth taking the time
and the risk of orienting even further in the wrong direction
which a tumble will involve.

\begin{figure}
\centerline{ \epsfxsize = 4in
\epsffile{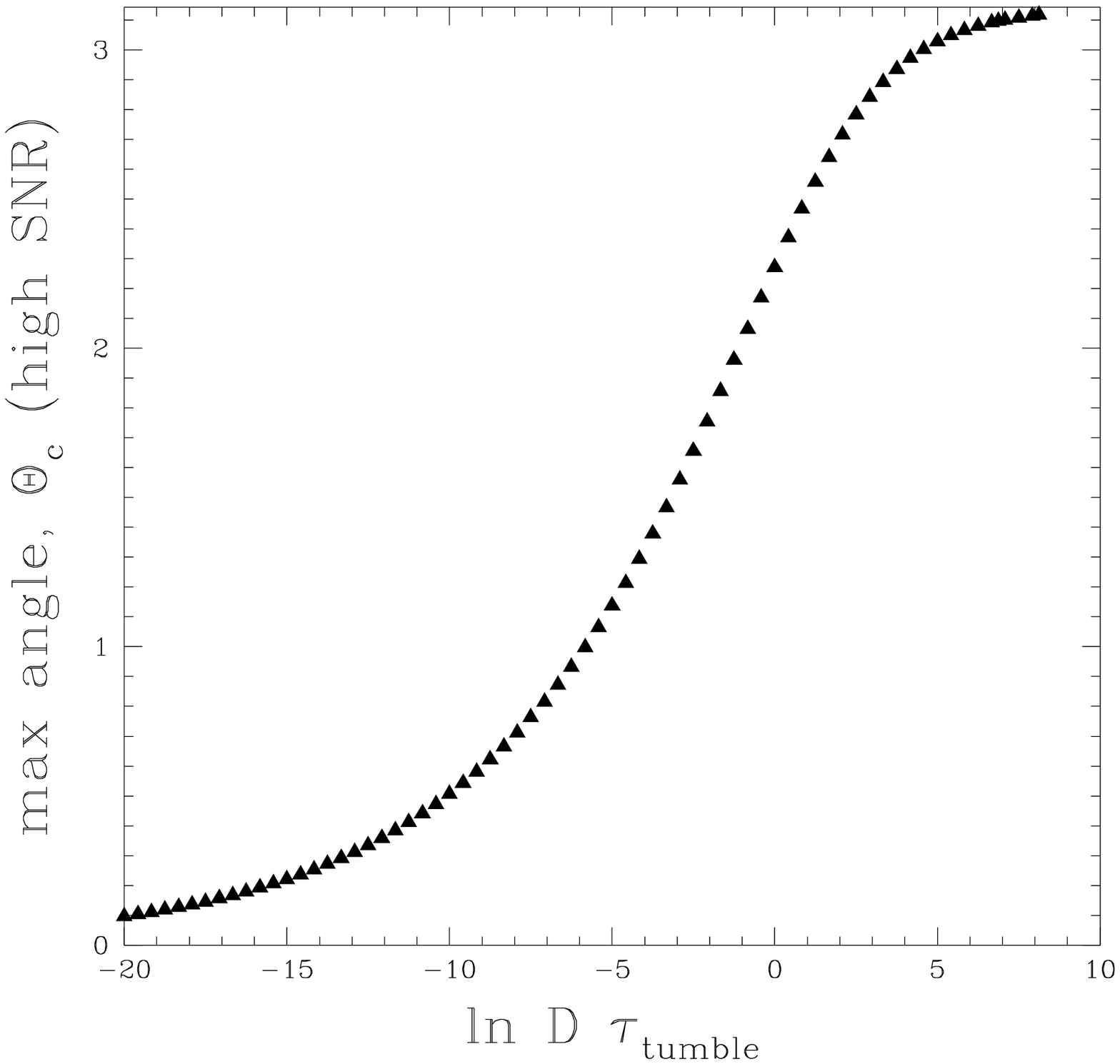}}
\caption{Plot of the optimal angle at which the bacterium
should initiate tumbles versus the product of its rotational
diffusion constant and the duration of the disorienting tumbles.}
\label{fig:angle}
\end{figure}

It is clear that the dimensionless
number $D_{\rm rot}\tau_{\rm tumble}$
fixes $\Theta_c$ and
one would expect that, for small values
of $D_{\rm rot}\tau_{\rm tumble}$,
$\Theta_c$ would also be small, vanishing
like $(D_{\rm rot}\tau_{\rm tumble})^p$.
In practice, 
$D_{\rm rot}\tau_{\rm tumble} 
\sim (.15 {\rm sec}^{-1})( .15 {\rm sec}) \sim .02$,
which 
is indeed small, and we might expect that
$\Theta_c$ would also be small. However we will see that this
is not the case as
$p = 1/6$ and the prefactor between
$(D_{\rm rot}\tau_{\rm tumble})^{1/6}$ and $\Theta_c$
is important.  The correct value of $\Theta_c$
can be obtained from the following argument:
the equilibrium distribution of orientation of 
running {\it E. Coli} for a
specific choice of $\Theta_c$
can be obtained from solving the heat equation
on the surface of a sphere
with a perfectly absorbing boundary condition
at $\Theta_c$ and a source term that deposits
new {\it E. Coli} uniformly over the allowed region
at a rate that exactly cancels the current
into the boundary at $\Theta_c$.  From the
distribution, one can compute
$\langle \vec{v} \cdot \frac{\nabla c}{|\nabla c|} \rangle$
by first computing the average over the 
distribution of running
{\it E. Coli} and then multiplying by the fraction of
{\it E. Coli} which are running. The latter is given
by the integral of the distribution over
the allowed region of the sphere divided
by the current into  the boundary of the allowed region
times the average length of time between runs.
The last factor is given by the average length of
a tumble times a factor,  
$(1-\frac{\rm forbidden~area}{4\pi})^{-1}$, which reflects the chance
for a tumble to result in an orientation
past $\Theta_c$, in which case it will
be followed
immediately by another tumble.
The relevant heat equation on the sphere reads:
\begin{equation}
\frac{\partial\rho}{\partial t} = C + 
\frac{D_{\rm rot}}{\sin\Theta}\frac{\partial}{\partial\Theta}
\left( \sin\Theta \frac{\partial\rho}{\partial\Theta} \right)
\end{equation}
$C$ can be fixed by
the requirement that $\rho$ vanish at $\Theta_c$.
The resulting, unnormalized solution is:
\begin{equation}
\rho(\Theta) = 
\ln\left(\frac{\cos\Theta/2}{\cos\Theta_c/2}\right)
\end{equation}
\begin{equation}
C = D_{\rm rot}/2
\end{equation}
Both $C$ and $\rho$ can be multiplied by an arbitrary,
common factor
since we have not yet imposed any normalization.
The ``current '' into the boundary
is given by
\begin{equation}
J_{\rm loss} = 2\pi C (1-\cos\Theta_c)
\end{equation}
while the integral of $\rho$ is given by
\begin{equation}
2 \pi \int_0^{\Theta_c} d(-\cos\Theta) \rho(\Theta) = 
\pi(\cos\Theta_c -1 -4 \ln\left(cos\frac{\Theta_c}{2}\right))
\end{equation}
The 
expectation value of $\vec{v} \cdot \frac{\nabla c}{|\nabla c|}$
is given by
\begin{equation}
\label{eq:return}
\langle \vec{v} \cdot \frac{\nabla c}{|\nabla c|} \rangle =
v |\nabla c| \frac{\sin^4 \frac{\Theta_c}{2} }
{ 2 D_{\rm rot}\tau_{\rm tumble} - 
2 ( \sin^2\frac{\Theta_c}{2} + 2 \ln\cos\frac{\Theta_c}{2}) }
\end{equation}
In general, the maximum can be found numerically,
but for the special case of
$D_{\rm rot} \tau$ small, 
where one expects small
$\Theta_c$, one can expand the 
trigonometric functions
to find for $x = \Theta_c/2$:
\begin{equation}
\langle \vec{v} \cdot \frac{\nabla c}{|\nabla c|} \rangle 
\sim
v |\nabla c| \frac{480 x^4 - 80 x^6 + 6 x^8 + \dots}{15360
D_{\rm rot}\tau_{\rm tumble} + 480 x^4 + x^8 + \dots}
\end{equation}
anticipating the result
that $D_{\rm rot}\tau_{\rm tumble} \propto x^6$
and expanding again
one finds that the derivative
vanishes for:
\begin{eqnarray}
x & \sim & \sqrt[6]{6 D_{\rm rot}\tau_{\rm tumble} } \\
\Theta_c & \sim  & 2 \sqrt[6]{6 D_{\rm rot}\tau_{\rm tumble} } \\
\end{eqnarray}
already for $D_{\rm rot}\tau_{\rm tumble} \sim .01$,
$\Theta_c$ is not small 
(the above formula is valid only if it is small
and would predict $\sim .4\pi$ already at this point).
The value of $\Theta_c$ obtained from 
numerically finding the value which maximizes
the return in depicted in Fig. \ref{fig:angle}.
It rises very rapidly to the vicinity of
$\pi/2$ where it remains for all reasonable values of
$D_{\rm rot}\tau_{\rm tumble}$, until ultimately
the behavior for very
large $D_{\rm rot}\tau_{\rm tumble}$
is given by
\begin{equation}
\Theta_c = \pi - \sqrt{ 2 /D_{\rm rot}\tau_{\rm tumble}}
\end{equation}
We see that, 
under a broad range of circumstances, the optimal strategy
is essentially to continue if one is moving into
regions of higher concentrations of attractant
and to reorient if one is moving into regions
of lower concentration of attractant. 
The bacterium
should be surprisingly tolerant of moving in the directions
that are far from perfectly aligned with
$\nabla c$,
even if it has perfect information as to how far off course it is.
In fact, the fraction of time the bacterium spends tumbling, $f$, is
not $\sim 1$, as one would naively expect,
but rather is given by:
\begin{eqnarray}
f & = & \frac{\tau_{\rm tumble} J_{\rm loss} }
{ (1-\frac{\rm forbidden~area}{4\pi})
\int 2\pi d(-\cos \Theta) ~\rho(\Theta) +
\tau_{\rm tumble} J_{\rm loss} } \\
& = &	\frac{2 D_{\rm rot} \tau_{\rm tumble} }
{ \cos \Theta_c -1 - 4 \ln\left(\cos \frac{\Theta_c}{2} \right) +
2 D_{\rm rot} \tau_{\rm tumble}} \\
& \sim &
\frac{5 D_{\rm rot} \tau_{\rm tumble}}
{1 + 5 D_{\rm rot} \tau_{\rm tumble}}
\\
& \sim & .1
\end{eqnarray} 
where we have used $\Theta_c \sim \pi/2$ and
$D_{\rm rot} \tau_{\rm tumble} \sim .02 $.
High frequency noise added to the thresholded quantity
will make the bacterium even more tolerant of
a signal indicating that it is swimming the wrong direction.

Before discussing noise, we first mention what strategy is
appropriate if the bacterium is,
for some reason, such as being in 
a spatially non-uniform gradient,
unable to estimate the magnitude of the
gradient and therefore the absolute
angle between its present orientation
and the gradient.  First, note that,
for the appropriate value of $D_{\rm rot}\tau_{\rm tumble}$,
the critical angle is close to $\pi/2$
so that,
approximately, only the sign of $\vec{v} \cdot \nabla c$ 
is important and the ignorance about $|\nabla c|$ is
unimportant.  In so far as it is important,
the optimal strategy in the absence of information
about $|\nabla c|$, depends on the
{\it a priori} probability $P(|\nabla c|)$
and is therefore somewhat non-universal,
depending strongly on the
statistics of the bacterium's environment
(it should consequently also be highly adaptive).
In particular, the optimal strategy is
now to 
choose a value of
$s_{\rm critical} = \vec{v} \cdot \nabla c$ at which to tumble
and the optimal value can be chosen
by maximizing:
\begin{equation}
\langle \vec{v} \cdot \frac{\nabla c}{|\nabla c|} \rangle 
\sim
\int d(|\nabla c|) P(|\nabla c|)
R\left(\Theta_c = \arccos (\frac{s_{\rm critical} }{|\nabla c|} ) \right)
\end{equation}
where
$R(\Theta_c)$ is defined by the right hand side of
Eq. \ref{eq:return}. The quantity to be maximized
corresponds to the average of the previously
calculated  
return over the 
actual
strategies that will result from a given
choice $s_{\rm critical}$. Clearly,
as $P(|\nabla c|)$ becomes sharply peaked,
this reduces to the case where
$|\nabla c|$ may be taken to be known,
in which case, as we have seen,
$s_{\rm critical} \sim 0$;
when the uncertainty in $|\nabla c|$
is large,
the bacterium should select an
even smaller $s_{\rm critical}$. 
The exact details depend on 
$P(|\nabla c|)$ in the environment to
which the bacterium is adapted.

Let us now discuss, 
in the case of a sharply peaked
$P(|\nabla c|)$, the introduction 
a small amount of high
frequency, Gaussian
noise to the bacterium's information about $\Theta$.
By the noise being small we mean that
its magnitude is much less than that
of the signal $v |\nabla c|$,
and
by high frequency we
mean that 
the characteristic time 
for $\vec{v} \cdot \nabla c$ to
to vary due to diffusion
$\sim  \Theta_c^2/ D_{\rm rot} $
is much larger than  
the noise time constant,
$\tau_{\rm noise}$,
defined by
\begin{equation}
\label{eq:noisetime}
\tau_{\rm noise} =
\sqrt{
\frac{\int d\omega~N(\omega)}{\int d\omega~\omega^2 N(\omega)}
}
\end{equation}
Here $N(\omega)$ is the power spectrum
of the noise so that
the mean squared magnitude 
of the noise, $\langle \eta^2 \rangle$,
is given by
\begin{equation}
\langle \eta^2 \rangle = \int \frac{d\omega}{2\pi} N(\omega)
\end{equation}
Such noise might arise inside the cell,
involving 
the thresholding mechanism
itself or the phosphorylation cascade,
or it might come from the
fluctuations in receptor occupancy
if the cell does not completely low
pass filter the receptor inputs.
To compensate for
the noise the optimal
strategy for the bacterium is to set its threshold,
$T$,
on the input, $\vec{v} \cdot \nabla c + \eta$,
where $\eta$ is the noise,
slightly lower
than in the noiseless case where
without noise $T = v \cos \Theta_c |\nabla c|$.
To determine $T$ note that
the rate at which
$\vec{v} \cdot \frac{\nabla c}{|\nabla c|} + \eta$
crosses the threshold, $T$, in one
direction is given by \cite{Kac}
\begin{equation}
r(\Theta)  =  \frac{\tau_{\rm noise}^{-1}}{2\pi}
\exp( - \frac{(T  - v \cos \Theta |\nabla c|)^2}{2 \langle \eta^2 \rangle} )
\end{equation}
The ideal value of $T$ can then be computed
by converting our previous
heat equation with a perfectly
absorbing boundary into one with
heat absorption  at angle
$\Theta$ given by
$ \rho(\Theta) R(\Theta)$.
For $~v \cos \Theta |\nabla c| > T$,
$R(\Theta)$ is given by
$r(\Theta)$ and
for $~v \cos \Theta |\nabla c| < T$,
by $(2\pi\tau_{\rm noise})^{-1}$.
Treating the tumble
rate as heat absorption in this way 
is exact if
the tumbling is a modulated
Poisson process.  
This is true in the
limit $\tau_{\rm noise} \rightarrow 0$
(a sensible answer in this limit
also requires $\langle \eta^2\rangle \rightarrow 0$),
but  the treatment is approximate
for noise with a finite correlation time.
However, we are interested in the short correlation time limit
and the approximation is justified.
In this limit, the heat effectively can not
penetrate into the region where
\begin{equation}
(v \cos \Theta |\nabla c| -T)^2 < 2 \langle \eta^2 \rangle
\ln(D_{\rm rot} \tau_{\rm noise})
\end{equation}
equivalently 
\begin{equation}
\theta > \arccos\left(\frac{T}{v|\nabla c|}\right) -
\sqrt\frac{2\langle \eta^2 \rangle |\ln(2 \pi D \tau_{\rm noise})|}
{v^2|\nabla c|^2 - T^2}
\end{equation}
It is important to realize that,
at this point, the
tumbling rate is rapidly increasing
because the value of the argument of
the exponential,
$\sim |\ln(2\pi D \tau_{\rm noise})|$
for small $\tau_{\rm noise}$,
is large and so is its derivative with
respect to $\Theta$.
The point effectively
specifies the location of
an absorbing ``wall'', and, therefore,
we are back to the case where our
original analysis applies and
the optimal strategy
is one which chooses $T$
so that 
$\theta$ lies at the $\Theta_c$
of the noiseless problem; this can be done
by choosing the threshold to be at
$v\cos\Theta_c |\nabla c|- 
\sqrt\frac{2\langle \eta^2 \rangle |\ln(2 \pi D \tau_{\rm noise})|}
{v^2 |\nabla c|^2 - T^2}$.
The threshold has been moved further from
$v |\nabla c|$
because the noise
results in tumbles being triggered prematurely.
Thus, the bacterium should be even more tolerant of
an input signal which indicates that it is
swimming in the wrong direction than
we observed that it should be in
the noiseless case. 

At this point it is worth making a few remarks about
the statistics of the runs and
tumbles in the high signal to noise ratio
case. For simplicity, let us consider
the noiseless case with known $|\nabla c|$.
The bacteria
emerge from tumbles with
random orientations and for
some tumbles the orientation immediately after
a tumble is past the critical allowed
orientation and they immediately
tumble again. This will lead to a renormalization
of the effective tumble length in this model
from $\tau_{\rm tumble}$ to
$\tau_{\rm tumble} \frac{2}{1-\cos \Theta_c}$,
however, the tumble durations will continue to be 
purely exponentially distributed, if
they were initially
exponentially distributed,
as they are found to be experimentally.
Meanwhile, if we average over all the bacteria
which do not immediately tumble to find the
statistics of run duration, we find these are also
roughly exponentially distributed. To see this,
recall that the heat equation we solved
to get $\rho(\Theta)$ is of the form:
\begin{equation}
\frac{\partial \rho}{\partial t} - \hat{L} \rho = C
\end{equation}
where $\hat{L}$ is a linear operator and
we impose the boundary condition $\rho(\Theta>\Theta_c) = 0$.
This could have
been solved by solving for the 
orthonormal eigenfunctions,
$f_{\lambda}(\Theta)$,
of $L$ which vanish at $\Theta_c$,
then re-expressing the linear source term, $C$,
as a sum of these:
\begin{eqnarray}
C & = & \sum_{\lambda} c_{\lambda} f_{\lambda}(\Theta) \\
c_{\lambda} & = & \int_0^{\Theta_c} d(\cos\Theta) f_{\lambda}(\Theta) 
\end{eqnarray}
Then we have:
\begin{eqnarray}
\rho(\Theta) & = & \sum_{\lambda} a_{\lambda} f_{\lambda}(\Theta) \\
a_{\lambda} & = & - \frac{c_{\lambda}}{\lambda}
\end{eqnarray}
The distribution of intertumble intervals
is then proportional to
$\sum_\lambda c_{\lambda} e^{-\lambda t}$.
The eigenfunctions
for a general $\Theta_c$ are hypergeometric functions
$_2F_1(-\nu,\nu+1;1;\zeta)$ where
$\zeta = \frac{1}{2}(1- \cos \Theta)$
\cite{Jackson}
and solving the boundary condition is  
in general impossible. However, for the
special case $\Theta_c = \pi/2$,
the eigenfunctions are those  
Legendre polynomials, $P_n(\cos \Theta)$
which vanish at $\pi/2$.
The eigenvalues are  $D n(n+1)$, 
where any odd $n$ are allowed.
The smallest eigenvalue is $2D$, and
the tail is therefore of the form $e^{-2 D t}$. The
next highest eigenvalue is $12D$ and the
coefficient in front of it at $t=0$, $c_{12D}$, 
is $\frac{7}{48}$ of the contribution from
the $n=1$ term, so the approach to exponential
decay of the form $e^{-2 D t}$ is very rapid.
This is consistent with the fact that
the mean duration of a run is,
in general, 
given by
$\frac{\cos\Theta_c -1 - 4 \ln\cos\frac{\Theta_c}{2}}
{D_{\rm rot}(1-\cos\Theta_c)}$, which for $\Theta_c = \pi/2$
gives $\frac{2 \ln 2 - 1}{D} \sim .38 D^{-1}$.  This
is only slightly smaller than $\frac{1}{2} D^{-1}$ because of
the inclusion of the higher $n$ terms.  Roughly
half of the shift is accounted for already
by the inclusion of the $n=3$ term.

Although we do not
believe that high signal to noise ratio limit
applies to the experiments of \cite{Btracking,tempcomp1,tempcomp2},
we should point out that the exponential tails
observed for the run distribution 
is consistent with the
statistics of runs and tumbles observed there.
The fact that our procedure 
for ending runs is essentially of the
threshold crossing type does not imply that
the distributions of run durations
has a  power law tail. This
contradicts the
claims of \cite{tempcomp2}, in
which  power law
tails were found for the
threshold crossings {\it of a random walk}
and argued to be typical for
threshold crossing processes.
A random walk is a very special
case since there is 
no finite correlation time
for the displacement.  Threshold
crossing a variable with a
finite correlation time results in a
tail for intervals between crossings
that is generally exponential.
The decay constant is roughly given by
$\tau_{\rm correlations}^{-1}
\ln({\rm prob.~no~crossings~in~}\tau_{\rm correlations})$,
and only for divergent or vanishing
$\tau_{\rm correlations}$ should non-exponential
tails result.
Thus, there is not, in general, any {\it a priori}
conflict between response regulator models and
the statistics of runs and tumbles.

\subsection{Low Signal to Noise Ratio Strategy}
\label{sec:lSNRstrat}

How should the bacterium make use of the information
available to it in 
small signal to noise ratio limit?  As in the
high signal to noise ratio case, the optimal strategy
is a deterministic algorithm for generating tumbles
based on the chemoreceptor binding histories.  In this case,
however, the bacterium will not tumble at fixed
angle because it can not determine its orientation with
respect to the gradient accurately.  Instead, it
will tumble when the output of some filtering 
of the chemoreceptor histories
crosses
some threshold which indicates that it is headed 
sufficiently in
the wrong direction.  
The problem of the optimal
strategy is to determine 
the filter, $F(t)$ and
the threshold.   This problem will be soluble
for two different assumptions about the nature of the
tumbles.  First, we will treat the case of completely 
disorienting tumbles of finite duration ($\tau_{\rm tumble}$),
as for the high signal to noise ratio case, and then
the case of instantaneous tumbles which do not
completely disorient the bacteria.

Let us introduce some notation.
In the low signal to noise ratio limit,
the filtering should be linear
and we denote the filtered output by
$\nu(t)$, defined by:
\begin{equation}
\nu(t) = \int_0^{\infty}  dt^{'} F(t^{'})
\dot{c}(t-t^{'})
\label{eq:fdef}
\end{equation}
where $c(t)$ is instantaneous concentration
inferred from the
fraction of occupied receptors, $\o(t)$.
Defining the concentration at which
half of the chemoreceptors will bind
attractant as  $c_{1/2}$, we have:
\begin{equation}
\label{eq:cdef}
c(t) = \frac{\o(t)}{1-\o(t)} c_{1/2}
\end{equation}
It is convenient to regard $\dot{c}(t)$
as the sum of two terms:
\begin{equation}
\label{eq:splitc}
\dot{c}(t) = \vec{v}(t) \cdot \vec{\nabla} c + \dot{\eta}(t)
\end{equation}
and denote their contributions to $\nu(t)$ by
\begin{equation}
s(t) = \int_0^{\infty}  dt^{'} F(t^{'})
\vec{v(t)} \cdot \vec{\nabla} c
\label{eq:sdef}
\end{equation}
\begin{equation}
n(t) = \int_0^{\infty}  dt^{'} F(t^{'})
\dot{\eta}(t)
\label{eq:ndef}
\end{equation}
The $\eta$ term represents the fluctuations 
in the chemoreceptor occupancy associated with the
stochastic binding of attractant molecules to
receptors.

Now let us compute the ``return'',
$\langle \vec{v} \cdot \vec{\nabla} c \rangle$
to be maximized.  For this,
we anticipate the result that
the characteristic time constant
of the filtered ``noise'', $n(t)$,
is short compared to that of the
filtered ``signal'', $s(t)$ (see Appendix A). 
We take the ``noise'' to be Gaussian
since it is a linearly filtered version of
the noise in the chemoreceptors,
and is expected to be  approximately Gaussian
via the Central Limit Theorem
(there are many receptors each making small
contributions to the total noise).

In this case, the rate at which
the filtered output, $f(t)$, crosses the threshold, $T$,
is given by:
\begin{equation}
\label{eq:cross}
r(s) = \frac{\tau_{n}^{-1}}{2 \pi}
exp( - \frac{(T-s)^2}{2 \langle n \rangle^2} )
\end{equation}
with
$\tau_{n}$
defined by
\begin{equation}
\tau_{n} = \sqrt{\frac{ \langle n^2 \rangle}{\langle \dot{n}^2
\rangle}}
\end{equation}
The trajectory of a bacterium emerging from a tumble
at time $t=0$ is therefore weighted by:
\begin{equation}
w(t,\{s\}) =  e^{-\int_0^t dt^{'} r\left( s(t^{'})\right)} 
\end{equation}
The return on the strategy is then
given by:
\begin{equation}
\label{eq:lowSNRreturn}
\langle  \vec{v} \cdot \vec{\nabla} c   \rangle = 
\frac{ \langle \int dt^{'} \vec{v}(t^{'}) \cdot 
\vec{\nabla} c  ~~w(t^{'},\{s\}) \rangle}
{\tau_{\rm tumble} + \langle \int dt^{'} w(t^{'},\{s\}) \rangle}
\end{equation}
where the numerator is the mean distance swum
up the gradient of attractant in a run and the 
denominator is the mean time required for a run.

Let us begin with the computation of the denominator.
The typical length of a trajectory is given by:
\begin{equation}
\tau_{\rm run} = \langle \int dt^{'} w(t^{'},\{s\}) \rangle
\end{equation}
This can be computed to lowest order in the signal to 
noise ratio more or less straightforwardly;   
the only caveat is that while $s(t)$ is
formally proportional to the gradient of
concentration,
there is the possibility that the coefficient
of proportionality could diverge if the filter,
$F$, has a long tail.
In fact, $s(t)$ can be partitioned into two pieces:
one coming from the contribution to the integral
from
times in the not too distant past,
i.e. not much longer ago than the rotational
diffusion time 
($\int_0^{\sim \tau} dt^{'} ~F(t^{'}) \vec{v}(t - t^{'})
\cdot \nabla c$,
where $\tau = (2D_{\rm rot})^{-1} $)
and another piece coming 
from the remainder of the integral
($x \equiv \int_{\sim \tau}^{\infty} dt^{'} ~F(t^{'}) \vec{v}(t - t^{'})
\cdot \nabla c$.
Only
the second piece can avoid 
being small for small concentration
gradient, and only if $F(t)$ decays 
slowly at long times.  In this case,
this term yields
a contribution to $s(t)$ which is
effectively static and is 
Gaussianly distributed since it is
essentially the sum of many independent
contributions from $s(t)$
stretching far into the past
(recall that $\vec{v}(t)$ decorrelates on a
time scale that is of order $(2D_{\rm rot})^{-1}$).
The mean value of this contribution, $\langle x \rangle $,
is zero and 
\begin{eqnarray}
\langle x^2 \rangle & = &
\frac{\tau_{\rm run}}{\tau_{\rm run} + \tau_{\rm tumble}}
\langle \int dt^{'} \int dt^{''}
 F(t^{'}) F(t^{''}) \vec{v}(t- t^{'}) \cdot \nabla
c ~~\vec{v}(t - t^{''})\cdot \nabla c \rangle \\
& = &
\frac{\tau_{\rm run}}{\tau_{\rm run} + \tau_{\rm tumble}}
\frac{2 v^2 |\nabla c|^2}{3 \gamma}  \int dt^{'} F^2(t^{'}) 
\end{eqnarray}
where we have assumed:
\begin{equation}
\langle \vec{v}(t) \vec{v}(0)  \rangle \propto e^{ - \gamma |t|}
\end{equation}
and low signal to noise ratio.
For the case where each tumble completely
disorients the bacterium, $\gamma$ is given at zeroeth
order in the concentration gradient by:
\begin{equation}
\gamma = \tau_{\rm run}^{-1} + 2 D_{\rm rot}
\end{equation} 

This second contribution, $x$,
to $s(t)$ is can be thought of as a contribution 
to the noise since it is uncorrelated with the
present orientation of the bacterium.  In fact,
since it and the occupancy noise are both Gaussian,
it can be integrated out. Let us therefore 
switch to using $s$ to refer only
to the first part since that is the part which
actually does contain the signal.
Then we find that:
\begin{eqnarray}
\langle r(s) \rangle_{\rm x = second ~part ~of ~s}
& = & \int \frac{dx}{\sqrt{2\pi\langle s^2\rangle}}
\frac{\tau_{n}^{-1}}{2 \pi}
\exp( - \frac{x^2}{2 \langle s^2 \rangle}) 
\exp( - \frac{(T-s - x)^2}{2 \langle n^2 \rangle} ) \\
& = & 
\frac{\tau_{n'}^{-1}}{2 \pi}
exp( - \frac{(T-s)^2}{2 \langle n'^2 \rangle}  )
\end{eqnarray}
where $n'$ is a new effective noise whose statistics
are those of the original noise, $n$, plus a
static contribution, $x$.  Thus
\begin{eqnarray}
\langle n'^2 \rangle & = & \langle n^2 \rangle + \langle x^2 \rangle \\
\tau_{n'} & = & 
\sqrt{ \frac{\langle n'^2 \rangle}{\langle \dot{n}^2 \rangle}}
\end{eqnarray}
Hereafter, we switch to using $n$ to refer to the effective
noise rather than using $n'$.
We are now also in a position to state our
definition of the signal to noise ratio:
\begin{equation}
\label{eq:SNR}
{\rm SNR} = \frac{ 9 \langle s \vec{v} \cdot \nabla c \rangle^2}
{v^2 |\nabla c|^2 \langle n^2 \rangle}
\end{equation}
The result is implicitly a function of the filtering
scheme and the tumbling rate, however, one generally
expects the signal to noise ratio to be proportional
to $v^2 |\nabla c|^2 / \langle \eta \eta \rangle $.
In fact, anticipating our
results for the optimal filter and tumbling rate we find that
in the low signal to noise ratio limit:
\begin{equation}
{\rm SNR} = \frac{ v^2 |\nabla c|^2 (1 + \bar{r} \tau_{\rm tumble})}
{N (8 D_{\rm rot} + 4 D_{\rm rot} \bar{r} \tau_{\rm tumble} - \bar{r} -
3 \bar{r}^2 \tau_{\rm tumble})(2D_{\rm rot} +  \bar{r})^2}
\end{equation}
where
\begin{eqnarray}
\langle \eta(t) \eta(t^{'}) \rangle & = &
\frac{N}{2 \tau_{\rm bare}} exp( - |t - t^{'}|/\tau_{\rm bare}) \\
&  \sim & N \delta(t - t^{'})
\end{eqnarray}

We now consider the return at lowest order in
the signal to noise ratio.
To zeroeth order in the concentration gradient, $r$ is constant over a run
and given by:
\begin{equation}
\label{eq:barr}
\bar{r} = \frac{\tau_{n}^{-1}}{2 \pi}
exp( - \frac{T^2}{2 \langle n^2 \rangle})
\end{equation}
and the mean run length is just $\bar{r}^{-1}$
and the denominator of the return expression,
Eq. \ref{eq:lowSNRreturn},
is just $\tau_{\rm tumble} + \bar{r}^{-1}$.

The numerator of
Eq. \ref{eq:lowSNRreturn},
can also be computed to lowest order in the
signal to noise ratio.
The redefinition of the ``signal'' and
the effective noise that we used for
calculating the mean run length is appropriate
here also so we continue to use $s$ to refer only
to the contribution from the recent past.
We expand the $w$ appearing in the
numerator to lowest order in $s$ to find:
\begin{eqnarray}
\langle \int dt^{'} \vec{v}(t^{'}) \cdot
\vec{\nabla} c  ~~w(t^{'},\{s\}) \rangle & \sim &  
v^2 |\nabla c|^2 
\int_{0}^{\infty} dt^{'} \langle \cos \Theta(t^{'})
\frac{ \partial }{ \partial (v \nabla c)}
\exp \left( -\int_0^{t'} r(t^{''}) dt^{''} \right)
\rangle \\ & \sim &
v^2 |\nabla c|^2 \frac{\partial \bar{r}}{\partial T}
\int_{0}^{\infty} dt^{'} 
e^{- \bar{r}t'}
\int_0^{t'} dt^{''} \int_0^{t''} dt^{'''}
F(t^{'''}) 
\langle 
\cos \Theta(t^{'}) \cos \left( \Theta(t^{''} - t^{'''}  \right)
\rangle  \\
& \sim &
 \frac{  (-T) v^2 |\nabla c|^2 }{ 3 
\langle n^2 \rangle  (2 D_{\rm rot} + \bar{r})}
\int_0^{\infty} F(t^{'''}) \exp( - (2 D_{\rm rot} + \bar{r}) t^{'''})
\end{eqnarray}
where we have used
\begin{eqnarray}
\frac{ \partial  \bar{r}}{\partial T} & = &
-\frac{\bar{r} T}{\langle n^2 \rangle}  
\end{eqnarray}

The return is therefore given by:
\begin{eqnarray}
{\rm return} & = &
 \frac{  \bar{r} (-T) v^2 |\nabla c|^2 }{ 3 
\langle n^2 \rangle  (2 D_{\rm rot} + \bar{r})
(1 + \bar{r} \tau_{\rm tumble})}
\int_0^{\infty} dt^{'''} F(t^{'''}) \exp( - (2 D_{\rm rot} + \bar{r}) t^{'''})
\\
& \equiv & -
\frac{ v^2 | \nabla c|^2}{3} \frac{T}{\langle n^2 \rangle} f(\bar{r})
\int_0^{\infty} dt^{'''} F(t^{'''}) \exp( - (2 D_{\rm rot} + \bar{r})
t^{'''})
\end{eqnarray}
To maximize the return we require that
$\frac{\partial {\rm return}}{\partial T} = 0$
and
$\frac{\delta {\rm return}}{\delta F(t)} = 0$.

For the $T$ equation we require 
\begin{eqnarray}
\label{eq:Tdef}
\frac{1}{T}
-\frac{1}{\langle n^2 \rangle}
\frac{ \partial  \langle n^2 \rangle}{\partial T}
+
\left(
\frac{\partial \ln f(r)}{\partial r}
- 
\frac{\int_0^{\infty} dt^{'''} t^{'''} F(t^{'''}) 
\exp( - (2 D_{\rm rot} + \bar{r}) t^{'''})}
{\int_0^{\infty} dt^{'''} F(t^{'''}) 
\exp( - (2 D_{\rm rot} + \bar{r}) t^{'''})}
\right)
\frac{ \partial  \bar{r}}{\partial T}
 & = & 0
\\
\frac{T^2}{\langle n^2 \rangle} =
\left(
\frac{2D_{\rm rot} - \bar{r}^2 \tau_{\rm tumble}}{
(1+\bar{r} \tau_{\rm tumble})(2 D_{\rm rot} + \bar{r})}
-
\frac{ \bar{r} \int_0^{\infty} dt^{'''} t^{'''} F(t^{'''})
\exp( - (2 D_{\rm rot} + \bar{r}) t^{'''})}
{\int_0^{\infty} dt^{'''} F(t^{'''}) 
\exp( - (2 D_{\rm rot} + \bar{r}) t^{'''})}
\right)^{-1}
\end{eqnarray}
where we have neglected subleading terms in the
signal to noise ratio.

Now we need to set up the filter equation 
and solve the two simultaneously.
For the filter equation we need to know:
\begin{eqnarray}
\frac{ \delta \bar{r}}{\delta F(t)} & = &
\frac{1}{2}
\left( \frac{\bar{r}}{\langle n^2 \rangle} \right)
\left(
\frac{T^2 - \langle n^2 \rangle }{\langle n^2 \rangle} 
 \frac{ \delta \langle n^2 \rangle }{\delta F(t)}
 -  
\tau_n^2
\frac{ \delta \langle \dot{n} ^2 \rangle }{\delta F(t)} 
\right)
\end{eqnarray}
We will also need to know
$\frac{\delta \langle n^2 \rangle}{\delta F(t)}$
and $\frac{\delta \langle \dot{n}^2 \rangle}{\delta F(t)}$,
for which we need to specify the
nature of the bare noise, $\eta$.
We take $\langle \eta \rangle = 0$
and 
\begin{eqnarray}
\langle \eta(t) \eta(t^{'}) \rangle & = &
\frac{N}{2 \tau_{\rm bare}} exp( - |t - t^{'}|/\tau_{\rm bare}) \\
&  \sim & N \delta(t - t^{'})
\end{eqnarray}
since the bare noise time constant is much
shorter than any other time scale in
the problem (see Appendix A).
In this case,
\begin{eqnarray}
\frac{\delta \langle n^2 \rangle}{\delta F(t)} & \sim &
- 2 N \ddot{F}(t) + \frac{4 v^2 |\nabla c|^2}{3 
(1+\bar{r} \tau_{\rm tumble})(2 D_{\rm rot} + \bar{r})}
F(t) 
\\
\frac{\delta \langle \dot{n}^2 \rangle}{\delta F(t)} & \sim  &
2 N \frac{\partial^3}{\partial t^3} {F}(t)
\end{eqnarray}
where these equations are valid for
regions where $F$ is slowly varying compared to
the bare noise time
and to
leading order in the signal to
noise ratio. 
The filter equation is given by
\begin{eqnarray}
0 & = & 
\left( 
\frac{\partial ln f(r)}{\partial \bar{r}}
-
\frac{\int_0^{\infty} dt^{'''} t^{'''} F(t^{'''})
\exp( - (2 D_{\rm rot} + \bar{r}) t^{'''})}
{\int_0^{\infty} dt^{'''} F(t^{'''})
\exp( - (2 D_{\rm rot} + \bar{r}) t^{'''})}
\right)
\frac{\delta \bar{r}}{\delta F(t)} \\
\nonumber
&  & -
\frac{1}{\langle n^2 \rangle}
\frac{\delta \langle n^2 \rangle}{\delta F(t)}
+ \frac{\exp\left( - (2 D_{\rm rot} + \bar{r}) t \right)}
{\int_0^{\infty} F(t^{'''}) \exp \left( - (2 D_{\rm rot} + \bar{r})
t^{'''} \right) } \\
\frac{\exp\left( - (2 D_{\rm rot} + \bar{r}) t \right)}
{\int_0^{\infty} F(t^{'''}) \exp \left( - (2 D_{\rm rot} + \bar{r})
t^{'''} \right) }
& = & 
\frac{1}{\langle n^2 \rangle}
\frac{\delta \langle n^2 \rangle}{\delta F(t)}
-
\left( \frac{\langle n^2\rangle}{\bar{r} T^2} \right)
\frac{\delta \bar{r}}{\delta F(t)}
\label{eq:filter} 
\end{eqnarray}
First, consider the behavior of the filter for times
large compared to
$(2 D_{\rm rot} + \bar{r})^{-1}$. In this
case the left hand side of Eq. \ref{eq:filter}
is negligible.  Further, the contribution to
$\frac{ \delta \bar{r}}{\delta F(t)}$
from the 
$\frac{ \delta \langle \dot{n}^2 \rangle }{\delta F(t)}$
is small, provided that $F$ is slowly varying in this
region.
So the filter equation requires,
\begin{eqnarray}
\frac{ \delta \langle n^2 \rangle }{\delta F(t)}  & = & 0 \\
\ddot{F}(t) & = & \frac{2 v^2 |\nabla c|^2 }
{3 N(1+\bar{r} \tau_{\rm tumble})(2 D_{\rm rot} + \bar{r})} F(t)
\end{eqnarray}
The absolute
scale of the filter is arbitrary given the definition
of $T$ used in Eq. \ref{eq:Tdef}, so
we may take:
\begin{eqnarray}
F(t) & \sim  & \exp  \left( - \sqrt{\frac{2 v^2 |\nabla c|^2 }
{3 N(1+\bar{r} \tau_{\rm tumble})(2 D_{\rm rot} + \bar{r})}} ~t \right)
\end{eqnarray}
The filter thus
has an extremely long tail determined
by the square root of the signal to noise ratio.

The behavior for intermediate times, times of
order $(2 D_{\rm rot} + \bar{r})^{-1}$,
is more complicated.  Now we must retain
the exponential term in Eq. \ref{eq:filter},
however, the term in $\frac{\delta \bar{r}}{\delta F}$
coming from
$\frac{ \delta \langle \dot{n} ^2 \rangle }{\delta F(t)}$
is still negligible and
we may also neglect the term in
$\frac{ \delta \langle n^2 \rangle }{\delta F(t)}$
that vanishes with the signal to noise ratio.
In this case we have:
\begin{eqnarray}
\left(
 \frac{1}{ 2\langle n^2\rangle }
+ \frac{1}{2 T^2 }
\right)
 \frac{ \delta \langle n^2 \rangle }{\delta F(t)}
 & = &
\frac{
\exp \left(- (2 D_{\rm rot} + \bar{r}) t \right)}
{\int_0^{\infty} dt^{'''} F(t^{'''}) \exp \left( - (2 D_{\rm rot} + \bar{r})
t^{'''} \right) } 
\\
\exp \left( - (2 D_{\rm rot} + \bar{r}) t \right)
& = & 
\left( \frac{1}{\langle n^2\rangle} + \frac{1}{T^2 } \right)
N \ddot{F}(t)
{\int_0^{\infty}dt^{'''}  F(t^{'''}) \exp \left( - (2 D_{\rm rot} + \bar{r})
t^{'''} \right) } 
\label{eq:midTconsist}
\end{eqnarray}
This requires that
\begin{equation}
F(t) = A + B t + C \exp \left( -(2D_{\rm rot} + \bar{r}) t \right)
\end{equation}
Matching the filter onto the result for long times
requires that
$A = 1$ and  $B=0$, while 
$F(0)= 0 $ requires 
$C = -1$  unless $F$ were to vary very rapidly in the
short time region, which is clearly not optimal 
because of the additional contribution to
the noise that would result.  In this
case,
$C= -1$ imposes a self-consistency
condition. 
Eq. \ref{eq:midTconsist} requires that
\begin{equation}
\left( \frac{1}{\langle n^2\rangle} + \frac{1}{T^2 } \right)
= \frac{2}{N (2D_{\rm rot} + \bar{r}) }
\end{equation}
where we have assumed that the very short time contribution
to $\int_0^{\infty}dt^{'''}  F(t^{'''}) \exp \left( - (2 D_{\rm rot} +
\bar{r}) t^{'''} \right)$ is negligible.
The contribution to $\langle n^2 \rangle$
from the the intermediate and long time
parts of the filter is
$\frac{N (2D_{\rm rot} + \bar{r}) }{2}$
so that 
there must be a short time
contribution to $\langle n^2 \rangle$
given by;
\begin{equation}
\langle n^2 \rangle_{\rm short ~times} =
\frac{N^2 (2D_{\rm rot} + \bar{r})^2}{4 T^2 - 2 N (2D_{\rm rot} +
\bar{r}) }
\end{equation}

The short time behavior of the optimal,
low signal to noise ratio filter depends on the characteristics
of the bare noise. The short time
equation for the optimum filter is:
\begin{eqnarray}
\label{eq:shorttimeF}
2 \left( 
\int_0^{\infty}dt^{'''}  F(t^{'''}) \exp \left( - (2 D_{\rm rot} +
\bar{r}) t^{'''} \right) \right)^{-1} & = &
\left( \frac{1}{T^2} + \frac{1}{\langle n^2\rangle }\right) 
\frac{\delta \langle n^2 \rangle } {\delta F(t)}
+
\tau_n^2 \frac{1}{T^2} \frac{\delta \langle \dot{n}^2 \rangle }
{\delta F(t)} 
\end{eqnarray}
where
\begin{eqnarray}
\frac{\delta \langle n^2 \rangle } {\delta F(t)} & = &
\frac{2}{\tau_{\rm bare}} F(t) - \frac{1}{\tau_{\rm bare}^2}
\int_0^{\infty} dt^{'} F(t^{'}) \exp(-|t-t^{'}|/\tau_{\rm bare}) \\
\frac{\delta \langle \dot{n}^2 \rangle } {\delta F(t)} & = &
-\frac{2}{\tau_{\rm bare}^3} F(t) 
-\frac{2}{\tau_{\rm bare}} \ddot{F}(t)
\\ \nonumber & &
+\frac{1}{\tau_{\rm bare}^4}
\int_0^{\infty} dt^{'} F(t^{'}) \exp(-|t-t^{'}|/\tau_{\rm bare})
\end{eqnarray}
For small $\tau_{\rm bare}$, we may neglect the left hand side
of Eq. \ref{eq:shorttimeF},
and the equation is the solved for a filter of the form
\begin{equation}
F(t) \sim A \left( 1- \exp( - t / \tau_{\rm bare}) \right)
\end{equation}
where $A$ must be chosen so that
\begin{equation}
\label{eq:taun}
\tau_n = \tau_{\rm bare} \sqrt{ 1 + \frac{T^2}{\langle n^2 \rangle} }
\end{equation}
This requires:
\begin{equation}
A = 2 \tau_{\rm bare} (2 D_{\rm rot} + \bar{r}) 
\frac{N (2 D_{\rm rot} + \bar{r})}{ 2 T^2 - N (2 D_{\rm rot} + \bar{r})}
\end{equation}
This choice of $A$ must also fulfill the self-consistency equation
from intermediate times, Eq. \ref{eq:midTconsist}.
A straightforward calculation demonstrates that, to leading
order in the signal to noise ratio, both self-consistency
requirements are satisfied by this choice of $A$ in the limit
of small $\tau_{\rm bare}$.
We thus arrive at the final form of the filter:
\begin{eqnarray}
\label{eq:finalF}
F(t) & = & \left (1 + 2 \tau_{\rm bare} (2 D_{\rm rot} + \bar{r})
\frac{N (2 D_{\rm rot} + \bar{r})}{ 2 T^2 - N (2 D_{\rm rot} +
\bar{r})} \right) \exp \left( - \sqrt{\frac{2 v^2 |\nabla c|^2 }
{3 N(1+\bar{r} \tau_{\rm tumble})(2 D_{\rm rot} + \bar{r})}} ~t \right)
\\ \nonumber & &
- 2 \tau_{\rm bare} (2 D_{\rm rot} + \bar{r})
\frac{N (2 D_{\rm rot} + \bar{r})}{ 2 T^2 - N (2 D_{\rm rot} +
\bar{r})}
\exp \left( -\frac{t}
{\tau_{\rm bare} } \right) \\ \nonumber & &
- \exp \left( (2 D_{\rm rot} + \bar{r}) t \right)
\end{eqnarray}

Notice that, as expected based on Berg and Purcell's
original arguments \cite{BandPmainchemorecept},
the time scale of the filter is
set primarily by the rotational diffusion
time of the bacterium.  Our filter is also similar
to that found in \cite{Leonid},
where a different criterion for optimality
was used.  However, the long time behavior found
here is somewhat different from that of \cite{Leonid}
and from that expected based
on the arguments of \cite{BandPmainchemorecept}.
We find that, in the low signal
to noise ratio limit, a temporal filter extending for 
significantly longer than the rotational diffusion time
is optimal.  
For weak signal strengths some useful
information is gained from averaging in measurements
made much longer ago than the rotational diffusion time
scale, and thus the optimal filter includes these times.
This is not what one would have naively
expected based on the arguments of \cite{BandPmainchemorecept}.
A similar effect was found in \cite{Leonid}, where the tail of
of the optimal filter was found to extend to
infinitely long times, however, that result was obtained in
the strictly zero signal case, rather than in the
limit of small signal as for our filter.  We find that
it holds only for the case of strictly zero signal to noise ratio,
(either for our definition of optimality or that used in
\cite{Leonid}) and that, for finite signal to noise ratio,
the optimal filter involves an additional time scale,
$\sqrt{3N(1+\bar{r}\tau_{\rm tumble})(2D_{\rm rot} + \bar{r})/
(2 v^2 | \nabla c|^2 )}$,
depending on the rotational diffusion time {\em and}
the signal to noise ratio.  
The tail of the filter decays on this time scale, which 
diverges with the inverse of the square root
of the signal to noise ratio 
but is of the same order as the rotational diffusion time
for finite signal strengths (where the calculations considered here are
not strictly valid but should be qualitatively correct).
Our results therefore unify the conclusions of 
of \cite{BandPmainchemorecept} and \cite{Leonid}:
for moderate signal
strengths, Berg and Purell's argument that the
filtering time scale will be of order the rotational diffusion time
will be correct, but at low signal to noise ratio a new,
extremely long, averaging time scale is optimal.

To compare Eq. \ref{eq:finalF} to experimental
results on the filtering strategies used
by the bacteria we require the equations which
determine $\bar{r}$,$T$, $\langle n^2 \rangle$ and $\tau_n$.
These are given by Eq. \ref{eq:barr},
\begin{equation}
\label{eq:T}
T = \sqrt{ \langle n^2 \rangle 
\frac{ 2(1+\bar{r}\tau_{\rm tumble})(2D_{\rm rot}+\bar{r})}
{4 D_{\rm rot} - 3\bar{r} -5 \bar{r}^2 \tau_{\rm tumble}} }
\end{equation}
\begin{equation}
\label{eq:nsq}
\langle n^2 \rangle  = \left(
\frac{2}{N (2D_{\rm rot} + \bar{r})} - T^{-2} 
\right)^{-1}
\end{equation}
and Eq. \ref{eq:taun},
respectively.  We can combine Eq. \ref{eq:T} and \ref{eq:nsq}
to obtain
\begin{equation}
T = \sqrt{\frac{N}{2}(2D_{\rm rot}+\bar{r}) 
\frac{8 D_{\rm rot} + 4 D_{\rm rot} \bar{r} \tau_{\rm tumble}
- \bar{r} -3 \bar{r}^2 \tau_{\rm tumble}}
{4 D_{\rm rot} - 3\bar{r} -5 \bar{r}^2 \tau_{\rm tumble}} }
\end{equation}
\begin{equation}
\langle n^2 \rangle  = \frac{N}{4} 
\frac{8 D_{\rm rot} + 4 D_{\rm rot} \bar{r} \tau_{\rm tumble} -
\bar{r} -3 \bar{r}^2 \tau_{\rm tumble}}
{1 + \bar{r} \tau_{\rm tumble}}
\end{equation}
The remaining equations must be solved numerically
and for the typical values \cite{Btracking}:
\begin{eqnarray}
\tau_{\rm tumble} & \sim & .15~{\rm sec} \\
2 D_{\rm rot} & \sim & .3~{\rm sec}^{-1} \\
\tau_{\rm bare} & \sim & 1 {\rm msec}
\end{eqnarray}
the resulting $\bar{r}$ is:
\begin{equation}
\bar{r} \sim .17 {\rm sec}^{-1}
\end{equation}
This value depends only very weakly on 
the poorly known $\tau_{\rm bare}$ and
the experimentally measurable $\tau_{\rm tumble}$
\cite{logfootnote}.
It depends strongly on $D_{\rm rot}$
and also on the assumption that each tumble
totally disorients the bacterium. For this
reason it is worth considering the problem
of tumbles which are not perfectly disorienting.

The low signal to noise ratio is also solvable
for tumbles which do not
perfectly disorient the bacterium
in the limit of
vanishing $\tau_{\rm tumble}$.
If we define
\begin{equation}
z = 1 - \langle \hat{v}_{\rm before} \hat{v}_{\rm after} \rangle 
\end{equation}
then we find that the return (Eq. \ref{eq:return}) becomes:
\begin{equation}
{\rm return} = \frac{v^2 |\nabla c|^2}{3}
\frac{T}{\langle n^2 \rangle}
\frac{z \bar{r}}{2 D_{\rm rot} + z \bar{r}} 
\int_0^{\infty} dt F(t) \exp \left( - (2 D_{\rm rot} + \bar{r}) t \right)
\end{equation}
In this case, all of the arguments for the case of perfectly
disorienting tumbles go through as before except that: (1) the
return is scaled by $z$, (2) $\bar{r}$ must be replaced
everywhere by $z\bar{r}$ and (3) $\tau_{\rm tumble}$ is to be
set equal to zero everywhere.
Roughly, the resulting optimal value of
$\bar{r}$ will be $1/z$ larger.  Experimentally, it
appears that $z \sim 1/2$ \cite{Btracking} and for this value we find:
\begin{equation}
\bar{r} \sim .37 {\rm sec}^{-1}
\end{equation}

\subsection{Comparison with Experiments}

Many of the features we would expect in the
behavior of bacteria implementing the
optimal strategy are directly comparable
to the observations of
Ref. \cite{Btracking} and \cite{Btrackingtemporal}, where
the behavior of free swimming {\it E. Coli} in 
spatial and temporal gradients of various
chemoattractants, as well as in the absence of 
such gradients, was studied.  

In Ref. \cite{Btracking},
free swimming {\it E. Coli} in the absence of
gradients were found to have a distribution of
run times that was approximately exponential
with a time constant of about $.85$ sec.  The
distribution of run times could be made almost perfectly
exponential by rescaling the run times 
by the mean run times of the individual
bacteria.   This form for the distribution of run
times for an individual bacterium is in agreement with
the form we find in the low signal to noise ratio limit,
the relevant limit in this case.  The fact that the time
constant is different for different bacteria
is also a natural for the optimal strategy
because  the bacteria differ in
their rotational diffusion constants, and, therefore,
different bacteria should choose different 
rates for initiating tumbles.
It would be very useful to see if differences in 
the rotational diffusion constants 
of individual bacteria correlate with their different tumbling rates,
a question not investigated in \cite{Btracking}.
Assuming that the disorientation due to tumbling remains
fixed, the tumbling rate from the optimal low
signal to noise ratio strategy should be roughly proportional
to the rotational diffusion constant, if the bacteria
are pursuing the optimal, adaptive strategy.

Whatever the nature of the correlations between
rotational diffusion and rates of tumble initiation,
the mean rate of tumbling observed, $\sim 1.2$ sec$^{-1}$,
is anomalously large.  It is a factor of 3 larger
than the rate we would expect even if we take into account
the correlation between orientations before and after
tumbles reported in \cite{Btracking}.  It is possible that
there is significant error in either the value of
$D_{\rm rot}$ or of $z$ that we have  used and it would
be of great value to have precise experimental determinations
of these from tracking experiments, since they are both
experimentally directly measurable.  
However, a factor of 3 
appears to be too much to be the result of inaccuracies
in these values, and Ref. \cite{Btracking} quotes a mean
change in orientation from beginning to end of
run of only 23 degrees, implying that the bacteria really
do run for times significantly shorter than the time which
disorients them.   In the low signal noise ratio limit
this is not optimal; however, this behavior may be the result of the
experiments involving bacteria
{\it in the absence of any chemoattractant},
rather than merely in the absence of gradients.  In fact,
bacteria in a uniform solution of $10^{-4}$ molar serine 
have an exponential run distribution with a time constant
that is roughly three times longer than was found in the
absence of serine \cite{Btracking}.  This agrees rather well
with the value expected for the optimal strategy, however,
it should be noted that a uniform concentration of
aspartate, a different chemoattractant, was not found to have
the same effect.  Clearly it would be desirable to have
more tracking work done in uniform or nearly uniform
solutions of chemoattractant that
are as similar as possible to the natural environment of {\it E. Coli}
in order to settle this question.  This would enable
us to determine whether the tumbling rate is really
anomalously large compared to the optimal.

In addition to determining the distribution of
run times in the absence of chemoattractant,
the observations of \cite{Btracking} also 
measured some of the effects of small gradients
on this distribution.
One of the interesting things found was
an indication of a peculiar asymmetry in the response of
the bacteria to small gradients.  {\it E. Coli} tumbled less often
when swimming up the gradient of chemoattractant,
but not more often when swimming down the gradient of
chemoattractant.  We have seen that at high signal to
noise ratio, bacteria following the optimal
strategy will be surprisingly reluctant to tumble because
of the finite amount of time required to tumble and the
fact that there is some chance that rotational diffusion will
improve their prospects.   This may explain some of
the observed reluctance to tumble if the bacteria
are in a medium signal to noise ratio regime.  In fact,
the asymmetry is only clearly observed for runs that are longer
than $1.5$ sec, so perhaps a minimum integration time,
and the accompanying boost in signal to noise ratio,
is required for the asymmetry.  Any detailed attempt to
explain the asymmetry would require more detailed
measurements of the responses of
{\it E. Coli} than currently available for free swimming
bacteria, and this is another area where the collection
of more data on free swimming bacteria would be of 
great value.

At low signal to noise ratio, 
there is  a possibility for asymmetry 
in responses other than the cost of tumbles:
we expect an exponential, not linear,
dependence of the tumbling rate on 
the sensory input.  In fact, Ref. \cite{Btrackingtemporal}
observed the response of free swimming {\it E. Coli} to 
temporal gradients of  glutamate and found that
their results were best fit by an exponential dependence
on the rate of change of receptor occupancies, which
in the concentration region treated implied an
exponential sensitivity to gradients in concentration.  They did not,
however, claim to have ruled out a linear dependence 
and further measurements of the response of the
tumbling rate in bacteria adapted to a low signal
to noise ratio environment would be of great value.
It would also be of great value if bacteria could be
placed in a range of spatial gradients of chemoattractant
and then stimulated with an additional temporal
gradient to see if the response 
crossed over from one  appropriate for
the low signal to noise ratio case
to one appropriate for 
the high signal to noise ratio case, i.e.
tumbling at fixed value of the estimated angle from
the concentration gradient.

The predictions we have made for the optimal filter
to be used in the low signal to noise ratio limit
can be compared to the results of 
\cite{impresp,tempcomp1,tempcomp2},
where tethered \cite{tethernote} bacteria were exposed
to impulse like bursts of chemoattractant.
If the strategy the bacteria employ involves
linearly filtering some function of sensory inputs,
then,
the response a time $t$ later under these conditions
is related to $\dot{F}(t)$, and so the temporal properties
of the derivative of the filter can be determined from
these experiments.
Evidence for the very long tail in the 
derivative of the  filter 
expected of the optimal strategy for low signal to noise
was not found in these experiments. 
However, the filter found there does
extend about 4 sec into the past, which is significantly
longer than the time scale for the bacterium to
disorient. The improvement in chemotactic performance
that would result from a longer integration time is
very small, while the difficulty of building a faithful,
long term memory  is clearly substantial, so the
result is not surprising.   The long time behavior of the
filter may still be adaptive but the limit of
truly small signal to noise ratio appears to be 
beyond the range (if any) of that adaptation.

At intermediate times the derivative of the optimal filter
contains a term
with an exponential decay rate given by
$2D_{\rm rot} + z \bar{r} \sim 1 ~{\rm sec}^{-1}$
(for the realistic assumptions of short tumble times
and finite disorientation during tumbles). 
A feature of almost exactly those characteristics
is seen in the filter found in \cite{tempcomp1},
and it appears that the optimal
filtering strategy may in fact be rather
close to the filtering strategy inferred
from the response observed in those experiments.   
However, one should be cautious since the agreement
is the result of using the experimental value for
$\bar{r}$ which is not in agreement with theoretical
expectations (unless our value of $D_{\rm rot}$ 
or $z$ is
badly off), and there are worries about the
adaptive state of tethered bacteria \cite{tetherworry}.
Also, the tethered experiments were interpreted in terms
of a linear, but thresholded, response in the rate of
transitions from running to tumbling and vice versa; this
is not readily reconciled with the proposed deterministic
strategy and the resulting exponential dependence 
of the effective rate of such transitions on the
filtered signal.

It is probably not sensible to compare the
behavior of the filter at short times (100 msec or shorter)
with theoretical expectations since responses on
this time scale are not particularly important
for chemotactic performance, while constraints
due to the actual physical signal processing mechanisms
of the bacteria are severe on this time scale.

In summary, the filter observed has behavior very similar
to what is expected for the optimal filter at time scales
of order a second. On longer time scales it decays,
faster than optimal rate for low signal to noise
ratio, but at a rate that is clearly 
slower than the natural time scale for the bacterium
to disorient, in qualitative agreement with that 
characteristic of the optimal filter.  There is some
question as to the adaptation state of the tethered
bacteria, but since they spend most of their time in 
the complete absence of chemoattractant signals,
the most reasonable proposal is that they are adapted
to low signal to noise ratio.  The tethered experiments
do not make any attempt to identify adaptation to
signal to noise ratio or diffusion constant
in the characteristics of the filter.  In fact, they are ill
suited to such a test since the tethered bacterium
experiences such unnatural conditions that its 
adaptive state is difficult to determine.

On the other hand, some experiments suitable for
testing the form of the filter on free swimming
bacteria have recently been performed using 
photoreleased chemoattractant and repellant
\cite{photocaged}.
In those experiments, bacteria  swimming 
freely in the absence of spatial gradients were exposed to
step like changes in concentration by photoreleasing 
caged chemoattractants and repellants at the location of
the bacteria.  This form of stimulation provides
a delta function in the derivative of the concentration
so that the response a time $t$ later is a measure
of the filter, $F(t)$, used in the low signal to noise
ratio environment.
These experiments 
find a memory time for the system
slightly longer than that of the tethered experiments,
around 5 sec.  This is marginally closer to 
the long time tail is expected for
the optimal filter.
On the other hand,
there is no indication of the intermediate time ($\sim 1$ sec)
feature found in 
the data of \cite{impresp}, although the data
in the case of \cite{photocaged} are somewhat noisy and the applied
stimulus is so large, inducing a tumble in 100 msec
with nearly unit probability,
that a feature with the expected rise time of $\sim 1$ sec
is probably not excluded.  

It would be very useful to have
more
photorelease-based studies of the response of free swimming 
bacteria aimed specifically at  measuring the  properties of the filter
on time scales of a fraction of a second to several seconds.
One would like to know the filter properties
as a function of the spatial gradients to which the bacteria
are pre-adapted, and, if possible, one would also like to
look for correlations between the time constants
of the filter used, the rotational diffusion constant
of the free swimming bacteria and their distribution
of run times.  Experiments of this type could answer
the important open questions 
of the adaptability of the form of the filter 
and the tumbling criterion directly.
This would be a very important step in determining
the extent to which {\it E. Coli} achieves the
optimal chemotactic strategy. We believe that
this would provide interesting and important
information about
the limits of the sophistication of sensory processing
in single celled organisms.

B. Freedman acknowledges support from the Department
of Defense through an NDSEG Graduate Fellowship during
time spent at Harvard University
and S. P. Strong acknowledges support from the Department
of Energy through grant DOE DE-FG02-90ER40542 during time
spent at the Institute for Advanced Study.

\newpage

\section{Appendix A}

The correlation time
for the chemoreceptors is expected to be of order
$10^{-3} $sec or shorter.
This time scale is results from several considerations.
First,
the typical binding time for
a chemoreceptor is  about $10^{-4}$ sec
\cite{BandPmainchemorecept}.
This will set the correlation time for the chemoreceptor
inputs if:
(1)
it is  a much larger time than
the time for attractant to
diffuse away from a receptor, and,
(2) there is
relatively little correlation
between receptors. To see that (1) is satisfied
we approximate the time to diffuse away from the
receptor (that is
the time to diffuse far enough away to have small chance
of recapture) as
$\frac{{\rm the~size~of~the~receptor}^2}{{\rm diffusion~constant~of~
attractant}}$
which is typically
$10^{-14} {\rm cm}^2/10^{-5} {\rm cm~sec}^{-1}   \sim 10^{-9} {\rm
sec}$,
vastly smaller than the rotational diffusion
time, even if we have underestimated the
distance out to which recapture is important by a
sizable factor.
Individual receptors therefore decorrelate
on a time scale given by the binding time.  What about the ensemble of
receptors?

To see that the ensemble of
receptors decorrelates on a time scale comparable to
the binding time,
recall the arguments of \cite{BandPmainchemorecept} regarding
the rate of capture of diffusing molecules by
a large number of small perfectly absorbing
sites on the surface of an impermeable sphere.
The inbound current for $N$ patches of linear size
$s$ on a sphere of size $a$ is
given by:
\begin{equation}
J = 4 \pi D c_{\infty} a
\frac{Ns}{Ns+\pi a}
\end{equation}
where $c_{\infty}$ is the concentration of
signaling chemical per cubic centimeter
at infinity.
In \cite{BandPmainchemorecept}, it was
emphasized that
this differs from the current to a perfectly
absorbing sphere of radius $a$ only by
a factor of $\frac{Ns}{Ns+\pi a}$,
which can approach one for reasonable choices
of $N$, $s$ and $a$.  Here, we note that this
occurs because the current
$4 \pi D c_{\infty} a \frac{Ns}{Ns+\pi a} $
differs from that for
$N$ independent disks only by a factor of
$\frac{\pi a}{Ns + \pi a}$, which is about
one-half for the parameters of \cite{BandPmainchemorecept}.
Clearly, the reason for the
reduction from the value for $N$ independent
receptors is
that some of the molecules absorbed by a given
receptor would have contacted others in
its absence. In fact, the average molecule
that contacts an absorbing site would make contact
with another binding site
with probability $\frac{Ns}{Ns+\pi a}$.
For receptors which bind and release
attractant, this implies that the
average molecule which binds to one
receptor will bind to
$1/(1 - \frac{Ns}{Ns+\pi a})= \frac{Ns + \pi a}{\pi a} ~\sim 2$ others,
so that,
defining the number of receptors
with attractor bound at time $t$
to be $x(t)$ and assuming
Poisson statistics:
\begin{eqnarray}
\int_0^T dt \langle x(t) x(0) \rangle - T \langle x\rangle^2
& = & \left(\frac{Ns + \pi a}{\pi a} \right)
\tau_{\rm bind} \langle x\rangle \\
& \sim  & 2~\tau_{\rm bind} \langle x\rangle 
\end{eqnarray}
If the receptors decorrelated
on a time scale, $\tau_{\rm decorr}$,
which was much longer than $\tau_{\rm bind}$,
then
\begin{eqnarray}
\int_0^T dt \langle x(t) x(0) \rangle - T \langle x\rangle^2
&
\sim &
e~\int_0^T dt ~\exp(-t/\tau_{\rm decorr})~
\langle x(\tau_{\rm bind}) x(0)\rangle
\\
& \sim & e \tau_{\rm decorr} ~\langle x(\tau_{\rm bind}) x(0)\rangle
\end{eqnarray}
implying
\begin{eqnarray}
\langle x(\tau_{\rm bind}) x(0)\rangle  & \sim &
\frac{2 \tau_{\rm bind}}{e \tau_{\rm decorr}} \\
& \ll & 1
\end{eqnarray}
and the correlations among the receptors
must be weak if their correlation
time is much larger than the binding time.
Conversely, if $\tau_{\rm decorr} \sim \tau_{\rm bind}$,
then the correlations need not be weak, but
the two time scales are comparable
so that $\tau_{\rm bind}$ is still the appropriate time scale.

In practice, the time to diffuse far enough away
from {\it all} receptors should
be roughly
$\frac{{\rm the~size~of~the~bacterium}^2}{{\rm diffusion~constant~of~
attractant}}$
which is typically
$10^{-8} {\rm cm}^2/10^{-5} {\rm cm~sec}^{-1}   \sim 10^{-3} {\rm
sec}$ or ten times as long
as the receptor binding time.
In this case, most of the correlations among chemoreceptor
inputs have decayed by $.1$ msec but some
weak correlation
persist out to
$1$ msec.
Both times are
still very short compared to the
relevant rotational diffusion time scale,
and
we can approximate the noise in the binding of
the chemoreceptors to
be white on the other time scales of interest.


%
%

\newpage
\vspace*{-0.6cm}

%
%


\begin{references}
\bibitem{lowRfootnote} This method of propulsion has probably
evolved because in the low Reynolds
number
(vanishing ratio of inertial to viscous
forces) regime where bacteria operate,
propulsion can only result form motions
which break time reversal symmetry,
such as the rotation of flagella.
\bibitem{tumblefootnote}  In practice, all
tumbles  are not identical in that they
do not all take the same amount of
time; in fact, the distribution of tumble times is
approximately exponential with a time constant of
$\tau_{\rm tumble}$. However,
as mentioned in the text, we are assuming there is no correlation
between the duration of the tumble and
the sensory inputs the bacterium received prior to tumbling.
\bibitem{tethernote} Much of the quantitative 
work on the chemotactic response has been done on
bacteria which have been reduced to a single
flagellum by sonication and then attached to
a glass surface by the single flagellum. In this 
case, the bacterium rotates, rather than the flagella,
and it is possible to observe the sense of the
rotation and infer something about
whether the bacterium is ``trying'' to run or
tumble.
\bibitem{logfootnote} The return also depends on the
logarithm  of the time constant of the filtered noise,
however, that time constant stays finite as the
signal to noise ration vanishes, so that
in the low signal to noise ratio limit the
opitmal filter maximizes simply the signal to 
noise ratio.
\end{references}
\end{document}